\documentclass[reprint,amsmath,amssymb,notitlepage,noeprint]{revtex4-2}
\UseRawInputEncoding
\usepackage{graphicx}% Include figure files
\usepackage{dcolumn}% Align table columns on decimal point
\usepackage{bm}% bold math
\usepackage{braket} 
\usepackage{xcolor}
\usepackage{wrapfig}
\usepackage{balance}

\begin{document}

\title{Direct high resolution resonant Raman scattering measurements of InAs quantum dot dynamic nuclear spin polarization states}

\author{Aaron M. Ross${}^{1,2,3}$, Allan S. Bracker${}^{4}$, Michael K. Yakes${}^{4}$, Daniel Gammon${}^{4}$, L.J. Sham${}^{5}$, Duncan G. Steel${}^{1,2,*}$}
\affiliation{%
 ${}^{1}$H.M. Randall Laboratory of Physics, University of Michigan, Ann Arbor, MI 48109, USA  \\
 ${}^{2}$Electrical Engineering and Computer Science, University of Michigan, Ann Arbor, MI 48109, USA \\
 ${}^{3}$ Dipartimento di Fisica, Politecnico di Milano, Piazza L. da Vinci 32, 20133 Milano, Italy \\
 ${}^{4}$Naval Research Laboratory, Washington D.C. 20375, USA \\
 ${}^{5}$Department of Physics, University of California San Diego, La Jolla, California 92093, USA
}%

\date{\today}% It is always \today, today,
             %  but any date may be explicitly specified
\begin{abstract}
We report on the direct measurement of the electron spin splitting and the accompanying nuclear Overhauser field, and thus the underlying nuclear spin polarization (NSP) and fluctuation bandwidth, in a single InAs quantum dot under resonant excitation conditions with unprecedented spectral resolution. The dot consists of 10$^4$-10$^5$ nuclei, and is electrically biased to quantum confine an additional single electron. The electron spin splitting is measured directly via resonant spin-flip single photon Raman scattering detected by superconducting nanowires to generate excitation-emission energy maps. The observed two-dimensional maps reveal an Overhauser field that has a non-linear dependence on excitation frequency. This study provides new insight into earlier reports of so-called avoidance and tracking, showing two distinct NSP responses directly by the addition of a emission energy axis. The data show that the polarization processes depend on which electron spin state is optically driven, with surprising differences in the polarization fluctuations for each case: in one case, a stabilized field characterized by a single-peaked distribution shifts monotonically with the laser excitation frequency resulting in a nearly constant optical interaction strength across a wide detuning range, while in the other case the previously reported avoidance behavior is actually the result of a nonlinear dependence on the laser excitation frequency near zero detuning leading to switching between two distinct mesoscopic nuclear spin states. The magnitude of the field, which is as large as 400 mT, is measured with sub-100 nuclear spin sensitivity. Stable and unstable points of the Overhauser field distribution are observed, resulting from the non-linear feedback loop in the electron-trion-nuclear system. Nuclear spin polarization state switching occurs between fields differing by 160 mT at least as fast as 25 ms. Control experiments indicate that the strain-induced quadrupolar interaction may explain the measured Overhauser fields.
\end{abstract}
\maketitle

\section{Introduction}
%%% introduction
The quantum-confined electron spin in the single InAs QD system has been identified as a candidate for an optically-accessible qubit, due not only to the pristine nature of the photons scattered by the trion \cite{Matthiesen2012,Schaibley2013a}, but also due to the strong interaction between the electron and the nanoscopic nuclear spin ensemble comprised of 10$^4$-10$^5$ nuclei, which may provide a useful resource as a quantum memory or for quantum sensing applications \cite{Xu2009,Sun2012a,Gangloff2018}. There exists an extensive body of research providing evidence that the nuclear spins of InAs QDs interact with the confined electron and heavy/light-hole via non-linear feedback loops, with sensitive dependence on experimental parameters including the excitation laser frequency, polarization, power and pulse width/repetition rate; the effects are collectively referred to as dynamic nuclear spin polarization (DNP) \cite{Overhauser1953,Lampel1968,Meier1984,Gammon1997,Urbaszek2013,Debus2014,Debus2014a,Debus2015}. Many research groups have sought to understand, control, and reduce the impact of DNP, since it leads to loss of electron spin coherence \cite{Xu2009,Sun2012a,Yang2012,Yang2013,Gangloff2018,Chow2016,Stockill2016,Onur2016,Ethier-Majcher2017}. Significantly, optically-controlled fluctuation quieting has been reported  \cite{Xu2009,Sun2012a, Chow2016}, leading to extended electron spin coherence times up to at least 1 $\mu$sec without the need for dynamical decoupling \cite{Viola1999,Khodjasteh2005}. Various microscopic mechanisms have been proposed to account for DNP-induced effects related to NSP build-up and decay \cite{Maletinsky2007}, depending on the external magnetic field strength and direction, optical excitation conditions, and QD morphology \cite{Meier1984,Abragam1961,Slichter1992,Gammon2001,Koudinov2004,Hogele2012a,Maletinsky2007,Urbaszek2013}.

Previous reports on optical spectroscopy of the dynamic NSP processes in QDs have predominantly focused on non-resonant excitation of the QD, and subsequent detection of the photoluminescence (PL) on a diffraction grating spectrometer/monochromator \cite{Braun2006,Tartakovskii2007,Krebs2008}. Measurement of the NSP is performed by analyzing the polarization of the emitted PL or by the difference in energy between Zeeman-split trion peaks; the resolution of these experiments is limited by the grating dispersion in the best case to around 2.5 $\mu$eV (600 MHz). More fundamentally, non-resonant excitation of the QD has proven inadequate to yield single photons with high indistinguishability, as required for future quantum networks \cite{Flagg2012, He2013, Muller2014, Huber2015}. Far fewer experiments have been performed under resonant excitation conditions \cite{Xu2009,Latta2009,Stanley2014, Fernandez2009, Sun2016}, and to the authors' knowledge only one study has measured the Overhauser field directly during excitation of the neutral exciton transition via PL measurements using a sample bias switching technique with 50 $\mu$eV (12 GHz) resolution \cite{Kloeffel2011}.

Here we report on experiments in which the resonant Rayleigh and Raman scattering from the electron-trion system of a single InAs QD is spectrally resolved to directly report on the collective nuclear spin state mediated by the Overhauser (OH) field acting on the electron. Using Raman scattering, two dimensional maps of the OH field distribution obtained by measurements of a shift from equilibrium of the electron state splitting are constructed as a function of detuning from the trion resonance. The results show a surprising qualitative difference of the evolution of the NSP depending on which electron spin state is optically driven during the experiment, especially in terms of the polarization fluctuations. In one case (``tracking''), a relatively quiescent NSP changes monotonically with increasing excitation frequency; the NSP is characterized by a single peaked distribution function where the center of the distribution follows the frequency of the excitation field. In the other case (``avoidance''), the nuclear polarization appears to be unstable, indicated by multi-modal nuclear spin distributions with apparent switching between different nuclear spin configurations. We show directly that negative and positive feedback loops acting on the OH field are associated with dynamic optical excitation of the $\ket{x-}$ and $\ket{x+}$ electron spin states, respectively. With unprecedented spectral resolution, our experimental findings confirm and expand upon previous theoretical work \cite{Yang2012,Yang2013,Onur2018} showing that these non-linear feedback loops in the electron-trion system can be manipulated by optical excitation in order to tune the NSP in a single QD down to the sub-100 nuclear spin level.   

\section{Sample and experimental details}
The sample under study consists of InAs quantum dots embedded in a diode heterostructure, grown via molecular beam epitaxy. The QDs are approximately 2.5 nm in height. The heterostructure is grown on a 500 $\mu$m n-doped GaAs wafer (Si, $\geq 1\times 10^{18}/cm^3$), and consists of a distributed Bragg reflector (DBR) mirror (10 periods of 69 nm GaAs and 82 nm AlAs), 96 nm of n-doped GaAs ($\sim 2\times 10^{18}/cm^3$), 40 nm of undoped GaAs, the QD layer, 66 nm of undoped GaAs, 10 nm of n-doped GaAs ($\sim 1.5\times 10^{18}/cm^3$), 20 nm of undoped GaAs, 40 nm of p-doped GaAs (Be, $\sim 3\times 10^{19}/cm^3$), and a top DBR consisting of 4 periods of GaAs and AlAs with the same thicknesses as the bottom DBR. Two indium electrical contacts are used to apply a bias to DC Stark shift the QD energy levels into the 1$e^-$ stability range. The contacts are made following wet etching to two different layers. The lower contact is made to the 96 nm n-doped GaAs layer below the QDs and is annealed, while the top contact is non-annealed and is connected to the 40 nm p-doped GaAs layer beneath the top DBR. The bottom DBR is n-doped (Si, $\sim 1\times 10^{18}/cm^3$), while the top DBR was undoped. The top and bottom DBR layers are not affected by the electric field produced across the diode structure; as a result, the sample does not suffer from charge fluctuations in the DBR layers.

The diode heterostructure is operated by adjusting the bias voltage either to the edge of the charge stable range for the trion such that electrons actively tunnel between the n-GaAs electron reservoir and the QD at rates estimated from the trion linewidth and optical pumping times to be between $\sim$100-500 MHz (the co-tunneling regime), or to the middle of the range, where the electron is trapped in the QD for a prolonged period of time greater than 1 $\mu$sec.

The QD sample was held in a superconducting liquid helium magnet cryostat at 5.5 K. An in-plane magnetic field (Voigt geometry) is applied, resulting in a Zeeman shift that splits the trion into four linearly-polarized resonances. The in-plane electron and heavy-hole g-factors are determined using PL and differential reflectivity measurements in the co-tunneling regime (Appendix \ref{gfactor}) to be $g_e^x = 0.431 \pm 0.004$ and $g_{hh}^x = -0.332 \pm 0.004$, resulting in Zeeman shifts of (6.00 $\pm$ 0.05) GHz/Tesla and (4.64 $\pm$ 0.05) GHz/Tesla for the electron and heavy-hole, respectively. 

The experiments reported in this paper are performed using high-resolution CW Ti:Saph tunable lasers (Coherent MBR, 50 kHz linewidth). The incident laser beams are focused by a 0.68 NA aspherical lens onto the sample  held in a superconducting magnet cryostat, and the reflected light is collected.

Resonance Rayleigh and Raman scattering experiments are performed by rejecting the excitation/re-pump beams collected in the reflection geometry with polarization analyzers and waveplates. The scattered photons and remaining excitation fields are spectrally filtered by transmission through a pressure-tuned etalon (FSR = 45 GHz, FWHM = 400 MHz), collected into a single-mode fiber and detected with superconducting nanowire detectors (Quantum Opus).

Two-dimensional excitation-emission maps are constructed by rapidly scanning the excitation laser across a given resonance, with the pressure-tuned etalon energy fixed. 25 ms integration time per data point is used, and a laser scan rate of approximately 200 kHz/ms was chosen. The etalon energy is then iterated over a wide energy range in order to determine the OH field response to the scanning laser. Thus, the same laser scanning experiment is performed repeatedly (number of etalon energy steps) in order to construct a single map.

\section{High-resolution two-field resonant Raman scattering}
\begin{figure}[t]
\includegraphics[width=0.5\textwidth]{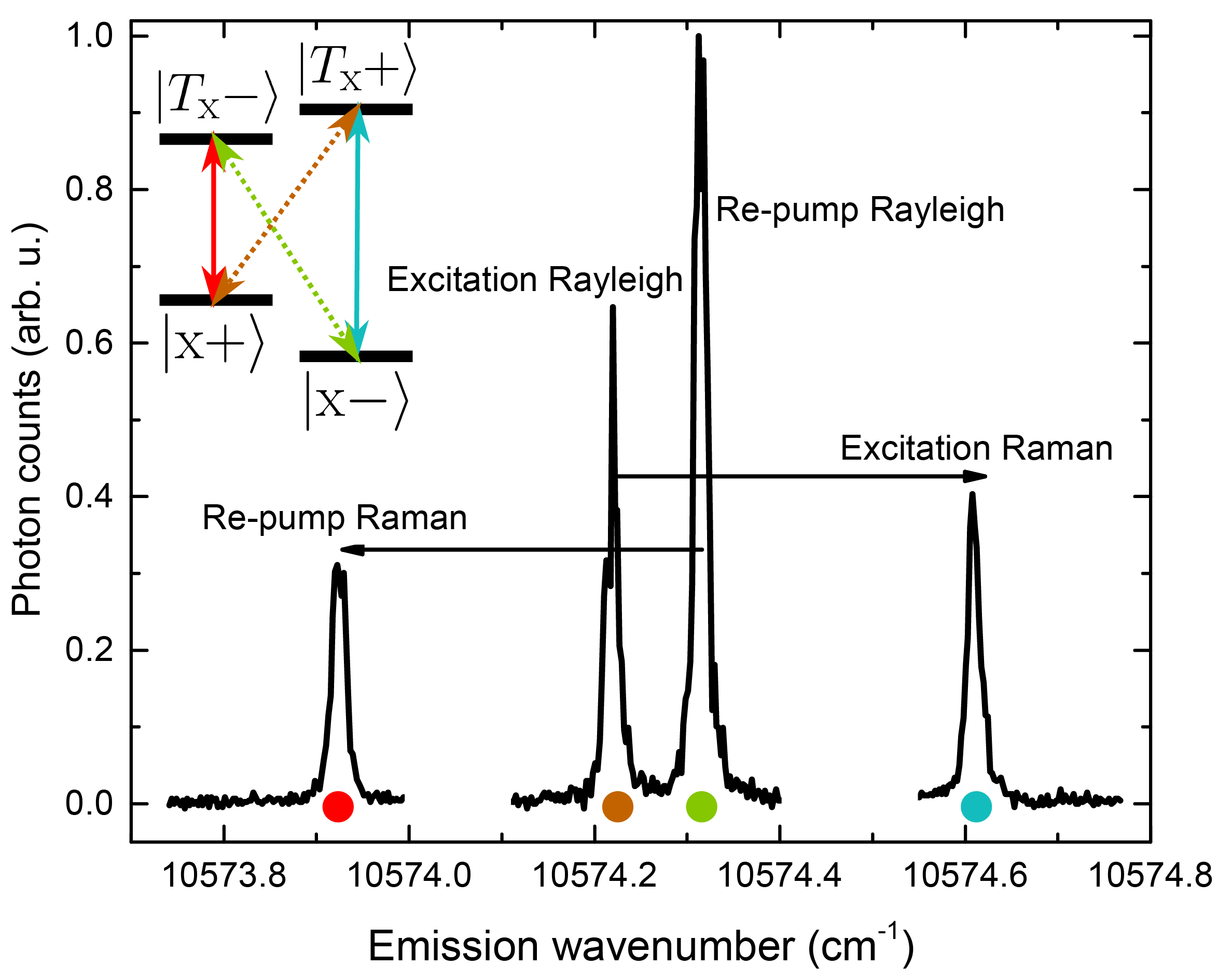}
  \caption{Resonant Rayleigh and Raman scattering in the optical pumping regime as measured by scanning a pressure-tuned etalon across all four scattering lines, with the excitation and re-pumping lasers held in resonance with the $\ket{x+}\rightarrow\ket{T_x+}$ and $\ket{x-}\rightarrow\ket{T_x-}$ transitions, respectively. Black arrows denote the relaxation pathways that correspond to the Stokes and anti-Stokes spin-flip Raman scattering lines for the re-pump and excitation lasers, respectively. Selection rules are denoted by solid (dashed) lines in the case of horizontal (vertical) polarization, where horizontal is along the magnetic field direction $\hat{x}$ and vertical is along the $\hat{y}$ direction generated by the right-handle rule with $\hat{z}$ pointing along the QD growth direction. Data was not collected in the blank regions between scattering lines.}
\label{fig:Figure1}
\end{figure}

The inset in Figure \ref{fig:Figure1} shows the 4-level energy diagram for the single electron in an InAs QD under the application of an in-plane 2 Tesla magnetic field (Voigt geometry). The ground and trion states in the Voigt geometry are defined as linear superpositions of the electron and heavy-hole states quantized along the growth axis z: $\ket{x\pm} \equiv \left(\ket{z-}\pm \ket{z+}/\sqrt{2}\right), \ket{T_x\pm} \equiv \left(\ket{T_z-}\pm \ket{T_z+}/\sqrt{2}\right)$. The heavy hole spins $\ket{T_z\pm} \equiv \ket{\pm\frac{3}{2}}$ are treated as an effective two-level system under strain \cite{Xu2007,Emary2007}. The resulting eigenstates are now quantized along the applied magnetic field direction. The electron spin state splitting is measured by using the pressure-tuned etalon to frequency resolve the Rayleigh and Raman scattering spectrum.  Fig. \ref{fig:Figure1} shows the basic spectrum at a sample bias where optical pumping occurs.  In the presence of optical pumping, which occurs in about 25 nsec (Appendix \ref{timedomain}, \cite{Xu2007}), a strong repumping field is utilized to enable scattering of multiple photons. In this experiment, this repumping is done by an optical field near resonance with the $\ket{x-}\rightarrow\ket{T_x-}$ transition, with the excitation field held in resonance with the $\ket{x+}\rightarrow\ket{T_x+}$ transitions.  A Rayleigh spectrum at the laser frequency associated with each optical field is seen along with the corresponding anti-Stokes and Stokes Raman spectrum from the excitation and repump field, respectively. The Raman shift, or the energy difference between the Raman and Rayleigh peaks for a given laser field are given by
\begin{equation}
    \Delta_{Raman} = E_\text{Raman} - E_\text{Rayleigh} = \pm \left( \Delta_e^{ext} + \Delta_{\text{OH}} \right)
\end{equation}
where $\Delta_e^{ext}, \Delta_{\text{OH}}$ are the external magnetic field splitting of the electron (0.402 cm$^{-1}$ (12.1 GHz) at 2 T) and the OH field, respectively, and the +(-) corresponds to an anti-Stokes (Stokes) scattered photon. To determine the OH field under bias conditions where the system is optically pumped (e.g, the data in Fig. \ref{fig:Figure1}), we measured the electron splitting due to the combined external and OH fields.  In the case of Fig. \ref{fig:Figure1}, this is measured to be 11.7 GHz (0.390 cm$^{-1}$). This differs from the 12.1 GHz measured under co-tunneling bias and resonant excitation, where no optical pumping occurs because of fast electron spin relaxation and minimal OH field is expected on resonance (e.g., Fig.\ref{fig:Figure2}). The difference is -400 MHz (0.013 cm$^{-1}$) showing that $\text{B}_{OH}$ = -66.3 mT. While not shown here, no significant differences in scattering lineshapes are observed between co-tunneling and optical pumping for stationary laser energies.

The OH field is connected to the optically-induced nanoscopic NSP in the QD by dividing the measured OH field shift by the maximum possible OH field shift in the absence of depolarizing mechanisms. In the simplest form, the maximum OH field in the InAs QD is given by $B_{OH}^{max} = \frac{1}{2\mu_B g_e}\left[ 3 A_{As} + 9A_{In} \right]$ = 13.2 T, where $A_{i}$ are the hyperfine constants for As, In \cite{Urbaszek2013}. In this experiment, the electron spin splitting, which is dynamically shifted only by the NSP via the OH field, is measured directly as the energy difference between the scanning excitation laser and the Raman scattered photons. The NSP is related to the electron spin splitting only by coefficients determined by the electronic wave function and atomic composition in the QD; therefore the qualitative response of the NSP to differing optical excitation conditions is also measured directly. 

We also note the linewidth of the Raman emission in Fig. \ref{fig:Figure1} is limited by the etalon linewidth (400 MHz, 0.013 cm$^{-1}$), which is smaller than the zero field reflectivity linewidth (545 MHz, 0.018 cm$^{-1}$), implying that there is minimal pure dephasing and spectral wandering of the electron spin.

Using the Raman emission from a single QD, we are then able to use this technique to generate two-dimensional maps of the OH field and hence determine the corresponding nuclear polarization as a function of the detuning of the optical excitation relative to the expected trion resonance.

\section{DNP revealed through excitation-emission Raman maps}
In the following sections, we investigate non-linear feedback loops resulting in NSP, referring to the action/back-action of the OH field (nuclear spin polarization) on the electron spin splitting, which leads to either reduced/increased optical excitation for a fixed laser excitation energy. Subsequent changes in electron/trion population lead to a change in the nuclear spin polarization, closing the feedback loop and resulting in feedback on the OH field and optical excitation \cite{Yang2012,Yang2013}. These feedback loops are measured via acquisition of two-dimensional excitation-emission Raman scattering maps for two cases: one in which the electron spin relaxation is rapid, referred to as co-tunneling, and the other for which the relaxation is extended. These two cases are differentiated by the bias condition where in the first case, co-tunneling prevents optical pumping and in the second case, co-tunneling is suppressed. In both cases, the dot remains negatively charged with 1 electron.    

\subsection{Co-tunneling regime}

\begin{figure*}[t]
  \includegraphics[width=0.85\textwidth]{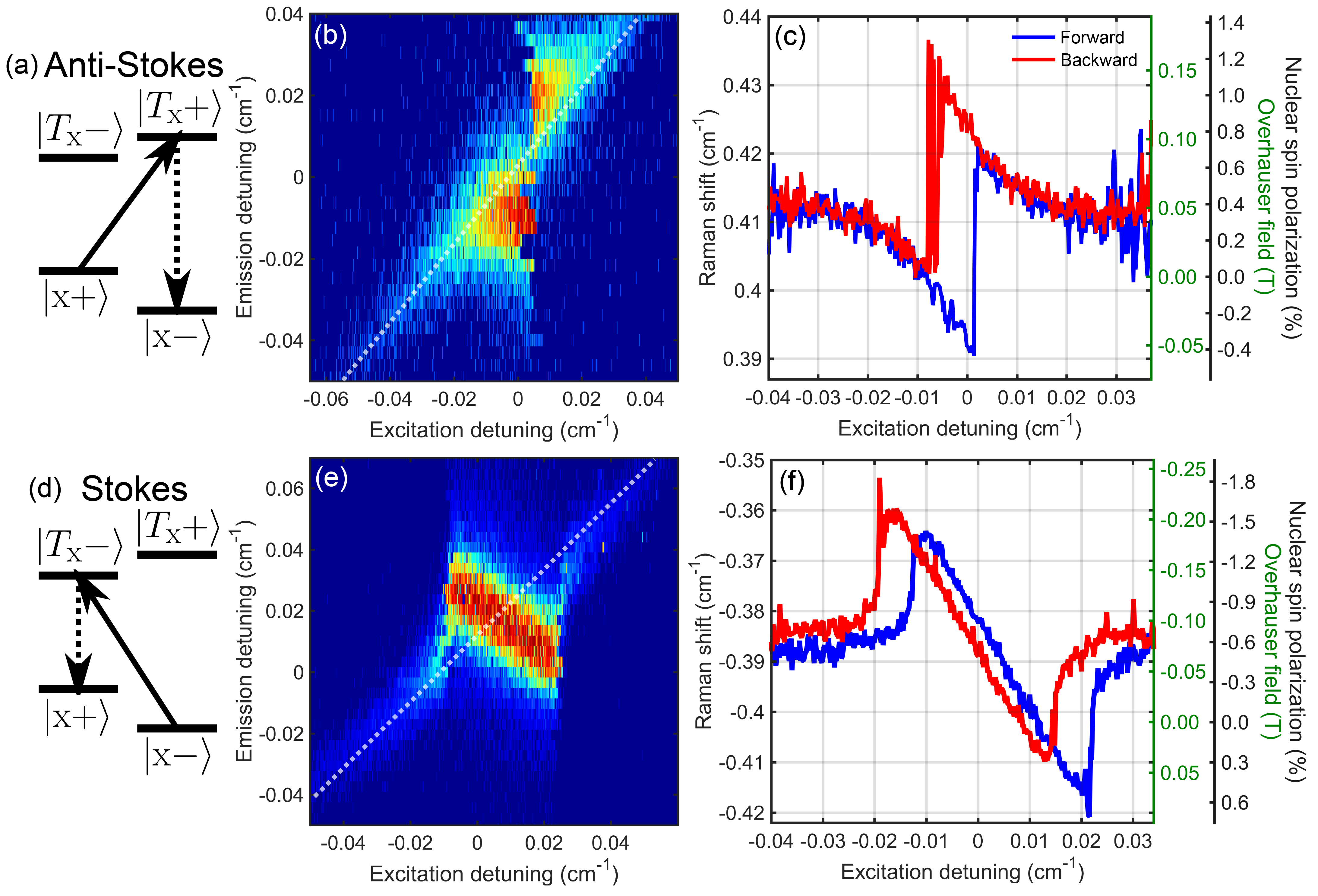}
  \centering
  \caption{2D Raman scattering excitation-emission energy maps and Raman shifts measured in the co-tunneling experiments. (a), (d): energy level diagrams for the anti-Stokes and Stokes scattering cases, with scanning excitation laser (Raman scattering pathway) indicated with an solid (dashed) arrow. (b), (e): 2D Raman scattering maps when the excitation laser is scanned across the $\ket{x+}\rightarrow\ket{T_x+}(\ket{x-}\rightarrow\ket{T_x-})$ transition. The excitation and emission detunings are measured relative to the extrapolated resonances calculated using the g-factors measured at co-tunneling and the Stark shift between co-tunneling and optical pumping. Dashed white lines indicate expected Raman scattering dependence in the absence of DNP. (c), (f): Raman shifts extracted by fitting each vertical cut of the 2D maps (b), (e) for a given excitation detuning, in both the increasing (blue) and decreasing (red) excitation laser energy directions. Axes on right-hand side of (c) and (f) indicate the OH field in Tesla, measured relative to the zero-DNP Raman shift equal to $\pm$ 0.402 cm${}^{-1}$ (green axis), and the NSP (blue axis). Non-linear color coding adjusted as described in Appendix D.}
\label{fig:Figure2}
\end{figure*}

Operating in the co-tunneling sample bias regime, electron spin relaxation is fast, preventing optical pumping. In this regime, there is no need for a strong re-pumping laser, resulting in a simplified spectroscopic study of the Raman scattering excitation-emission maps, revealing directly the NSP state (Figure \ref{fig:Figure2}).

In the first case, the excitation laser scanned across the $\ket{x+}\rightarrow\ket{T_x+}$ transition (Fig. \ref{fig:Figure2}a), yielding an anti-Stokes scattered Raman photon (Fig. \ref{fig:Figure2}b,c). In the absence of DNP, the center Raman scattering frequency is expected to track diagonally (white dashed lines in maps) as the laser is scanned. The Raman scattering maps and measured shifts in Fig. 2 are clearly anomalous: the Raman line is red shifted to avoid an apparently forbidden energy range, abruptly shifts blue by 0.03 cm$^{-1}$ (0.90 GHz, 150 mT, 1.14\% NSP), and pulls back asymptotically towards the expected diagonal line in the 2D map as a function of the detuning from the trion. The hysteresis between the increasing and decreasing excitation laser scan directions is pronounced near the expected zero-detuning region (Fig. \ref{fig:Figure2}c); the detunings at which the OH field switching occurs differ by 0.01 cm$^{-1}$ (300 MHz). 

Each excitation energy scan consistently displays either a sharp drop or rise that occurs on the order of 25 ms corresponding to the time to step the excitation laser 0.0002 cm$^{-1}$ (6 MHz), switching to an OH field which differs by $\pm$150 mT, corresponding to approximately $\pm$ 500 nuclei (assuming the QD consists of 5 $\times$ 10$^4$ nuclei), depending on the scan direction, demonstrating a remarkable sensitivity of both the DNP response to excitation conditions as well as the Raman emission to report on the nanoscopic nuclear spin configuration. 

Significantly, an identical experiment was performed in which the $\ket{x+}\rightarrow\ket{T_x-}$ transition was driven by the excitation laser to investigate the dependence of the DNP feedback on the heavy-hole spin projection: no discernible qualitative differences were observed (Appendix \ref{additional}). 

In the case described above corresponding to the excitation scheme in Fig. \ref{fig:Figure2}a, the OH field shifts the trion transition out of resonance with the laser, leading to an instability point in the vicinity of the expected trion transition as indicated by a sensitive switching of the OH field. The shift of the transition to an off-resonant condition has been called ‘avoidance’ \cite{Yang2013}. Prior to these experiments, these effects were observed indirectly by the apparent absence of the resonance in absorption \cite{Chow2016,Hogele2012a}. The experiments presented above are able to track the resonance and show that the \textit{NSP has shifted to reduce the optical excitation}.

In the second case (Fig. \ref{fig:Figure2}e,f), data is reported on the optical transition corresponding to excitation from $\ket{x-}$ to the $\ket{T_x-}$ state resulting in a Stokes-scattered Raman photon. The Raman scattering energy deviates considerably from the expected diagonal response in the 2D map at least as far away from the expected DNP-free resonance energy as 0.02 cm$^{-1}$ (600 MHz). The Raman energy is blue shifted over a short excitation range (0.005 cm$^{-1}$, 150 MHz) into a state that tracks in a nearly linear response as the excitation laser is scanned across the transition. The scattering intensity remains bright over an excitation range of 0.035 cm$^{-1}$ (1.04 GHz), until abruptly shifting blue again, at which point the scattered photon energy approaches the expected response far away from the expected resonance. This response shows that the nuclear polarization configuration is changing dynamically as the laser frequency is scanned. The OH field that reacts to the excitation laser is as large as -200 mT (1.51\% NSP). For comparison, for neutral excitons in GaAs fluctuation QDs under non-resonant conditions, excitation power-dependent NSP as large as 65\% has been reported with a polarization build-up time of around 3 seconds \cite{Gammon2001}. Additionally, there are two similar reports of so-called resonant ``dragging'' in InAs QDs with quantitative results. In the first example, the negative trion resonance is dragged by a resonant laser over 15 $\mu$eV \cite{Latta2009}, which corresponds to 520 mT (3.9\% NSP), assuming $g_e$ = -0.5; in the second case, the extracted OH field shift due to dragging is estimated to be around 4.1 $\mu$eV, or 140 mT (1.1\% NSP) \cite{Xu2009}. Thus, our findings are in general quantitative agreement with previous results involving resonant excitation, but provide direct measurements of the OH field and the corresponding fluctuations.

The hysteresis occurs primarily at the edges far away from resonance, where the system relaxes back towards an asymptotic approach parallel to the expected response curve, in stark contrast to the ``avoidance'' regime. Small offsets are observed between the expected and measured polarization responses in the forward and backwards scans at large detuning values throughout these experiments. As discussed later in this article, these small offsets may arise due to variations in the initial conditions of each experiment; sensitivities to initial conditions are expected in strongly non-linear systems. The corresponding experiment is also performed in which the $\ket{x-}\rightarrow\ket{T_x+}$ transition is probed: again, little difference is discerned between the two optically-excited heavy-hole projection states (see Appendix \ref{additional}). 

The DNP response described in the previous paragraph correlates with the so-called ``tracking regime'': the OH field that is generated by the optical excitation moves monotonically with the scanning excitation laser, acting to lock the trion resonance with the laser until a transition point is reached at which point the feedback mechanism breaks down and the nuclear polarization returns to its equilibrium value of zero.

The data here (Fig. \ref{fig:Figure2}c,f) shows the behavior of the OH field due to the nonlinear coupling between the NSP, the optical excitation of the electron spin, and the effects on the trion resonance. It is this non-linear behavior that gives rise to the distorted lineshapes first reported by Xu et al. \cite{Xu2009} and others \cite{Latta2009,Kloeffel2011,Hogele2012a}. The data provides a direct measurement of the shift of the electron spin energies and hence, the degree of the NSP. Additionally, changes in the electron spin splitting are measured independently of the trion energy, which is only related here to the intensity of the Raman scattering.  

We note the measurement approach reported here also does not require the addition of non-resonant optical excitation to extract the OH field, further simplifying the experiment and comparison with theory. As discussed above, the results differ from previous reports in terms of maximum NSP achieved, and the points at which hysteresis is observed to occur; we believe these differences result from the different experimental approaches taken, as well as inhomogeneity between samples and QDs. Furthermore, this technique allows us to study the fluctuations in the nuclear polarization distribution by examining the lineshapes of the Raman spectra as a function of excitation detuning under a variety of optical excitation conditions, as will be discussed later in this article.

\subsection{Optical pumping regime}

\begin{figure*}[t]
  \includegraphics[width=0.85\textwidth]{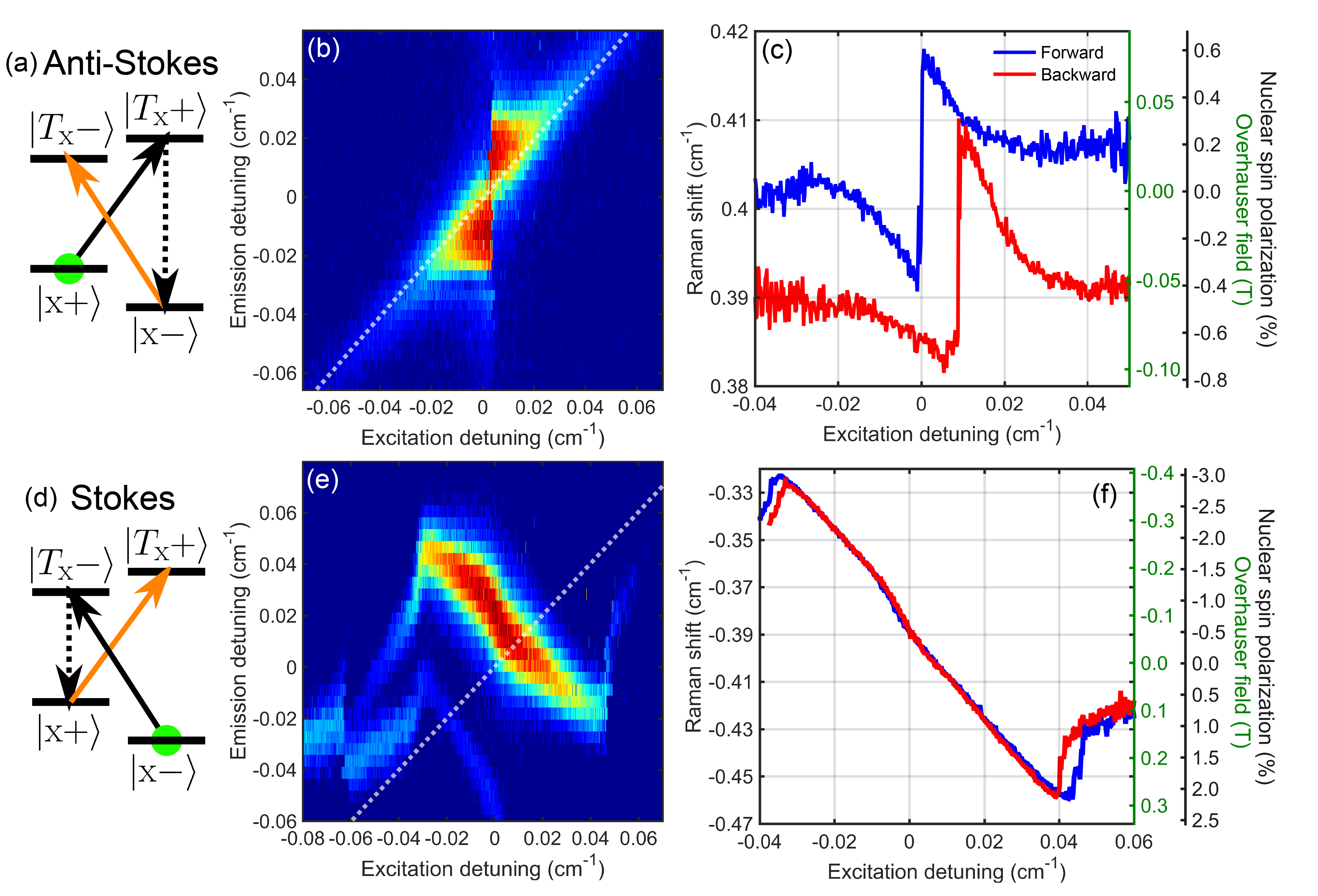}
  \centering
  \caption{2D Raman scattering excitation-emission energy maps and Raman shifts measured in the optical pumping experiments. (a), (c): energy level diagrams for the anti-Stokes and Stokes scattering cases, with scanning excitation laser/re-pump laser (Raman scattering pathway) indicated with a solid black/solid orange (dashed black) arrow, and the predominantly-prepared electron spin state indicated with a green circle. (b), (e): 2D Raman scattering maps corresponding to excitation schemes illustrated in (a) and (d), respectively. (c), (f): Raman shifts extracted by fitting each vertical cut of the 2D maps (b), (e) for a given excitation detuning, in both the increasing (blue) and decreasing (red) excitation laser energy directions.  Non-linear color coding adjusted as described in Appendix D.}
\label{fig:Figure3}
\end{figure*}

The next experiments are then performed when the QD is operated in a regime of minimal electron tunneling between the n-GaAs Fermi reservoir and the QD, referred to as the optical pumping regime \cite{Xu2007}. The electron is trapped in the QD at least as long as 1 $\mu$s (Appendix \ref{timedomain}), and the electron spin is depolarized at a much lower rate than in the co-tunneling regime. Operating in the optical pumping regime allows one to study how the nuclear spin ensemble evolves in the presence of a relatively pure electron spin state, as prepared by a strong optical re-pumping laser (1.5/10 excitation/re-pump power ratio). Additionally, the average 1e occupation factor is higher in the optical pumping regime than near the 0e and 2e co-tunneling edges; this factor has been demonstrated to dramatically affect the nuclear spin depolarization rate \cite{Wust2016a,Latta2011}. In the presence of optical pumping, emission (Rayleigh and Raman scattering) associated with the ground state of the excitation vanishes because the population of the ground state is depleted and transferred to the other ground state. Emission is restored by applying a second optical field to transfer (re-pump) the population back to the initial ground state.

In the first set of data taken in the optical pumping regime (Fig.\ref{fig:Figure3} a-c), a fixed re-pump laser (orange arrow) is held in resonance with the $\ket{x-}\rightarrow\ket{T_x-}$ transition, and the excitation laser is scanned across the $\ket{x+}\rightarrow\ket{T_x+}$ transition, driving the anti-Stokes Raman transition, similar to the first case in Fig. \ref{fig:Figure2}a-c. The electron spin state is prepared in a nearly pure $\ket{x+}$ state, with preparation fidelity estimated to be at least 94\% when the excitation laser is far off resonance (Appendix \ref{timedomain}). Measuring the anti-Stokes scattered photon energy as a function of excitation energy reveals that the DNP feedback responds in this case in an ``avoidance'' manner: when the excitation laser scans into resonance with the apparent trion transition, the OH field switches between states differing by an average 163 mT (1.23\% NSP), quantitatively comparable to the co-tunneling case. 

It is worth noting that the scans in which the laser excitation energy is increasing or decreasing differ in their baseline OH field by approximately 50 mT; repeated experiments indicate that this offset may depend on the extent to which the excitation laser is scanned off-resonance and the rate at which the laser is scanned. One conclusion is that the initial conditions of the electron-nuclear system affect the overall non-linear DNP feedback response, and that the system is more sensitive to these initial conditions in the optical pumping regime, but further experiments are required to understand this effect in more detail. The initial conditions include the NSP generated by the previous experiment; polarization decay times have been shown to be at least as long as seconds, if not minutes \cite{Maletinsky2007, Greilich2007, Sun2012a}. Nevertheless, the avoidance response to scanning the excitation laser across the $\ket{x+}\rightarrow\ket{T_x+}$ transition is common to both the co-tunneling and optical pumping regime. No differences are observed in the optical pumping regime when probing the $\ket{T_x-}$ optically excited state (Appendix \ref{additional}). 

In the second optical pumping experiment (Fig. \ref{fig:Figure3} d-f), the fixed re-pump laser drives the $\ket{x+}\rightarrow\ket{T_x+}$ transition, and the excitation laser is scanned across the $\ket{x-}\rightarrow\ket{T_x-}$ transition, similar to Fig. 2d-f, scattering a Stokes-shifted photon, thereby probing an initialized $\ket{x-}$ electron spin state. The qualitative responses are similar in the co-tunneling (Fig. \ref{fig:Figure2}) and optical pumping ``tracking'' (Fig. \ref{fig:Figure3}) regimes; however, the average OH field shift as measured from minimum to maximum is 2.5 times larger (724 mT) in the optical pumping regime than for co-tunneling, and the detuning range over which the tracking occurs is 2.4 times larger (2.49 GHz, 0.083 cm$^{-1}$). Another intriguing result is that the measured OH fields do not clearly return to the asymptotic expected dependence far away from resonance outside of the nearly linear tracking response, at least within the experimental range measured here. 

These results provide evidence that the magnitude of the optically-induced nuclear polarization in the optical pumping case for the $\ket{x-}$ Zeeman branch is considerably larger than in the $\ket{x+}$ Zeeman branch. This observation may possibly be understood as the result of two factors. First, the electron depolarization times are much longer in the optical pumping regime than in co-tunneling \cite{Latta2009,Wust2016a}; if the electron-nuclear coupling interaction depends on the magnitude of the electron spin polarization, then the resulting NSP may be enhanced \cite{Abragam1961}. Second, in the case where the laser is scanned across the transition involving the spin-down electron (e.g., Fig.\ref{fig:Figure3}d-f), the NSP is clearly locked to the excitation laser in a stable configuration. This stable configuration may reinforce the non-linear feedback mechanism, allowing for the build-up of larger NSP in the $\ket{x-}$ case. 

\section{Overhauser field distributions and dynamics}
OH field distributions can be constructed by taking emission energy slices of the excitation-emission maps for a fixed excitation laser detuning. These slices correspond to a high resolution measurement of the single photon Raman line shape. Fitting of the emission energy spectrum allows for determination of the OH field shift with very high precision. The 95\% confidence interval of the resonance center for the line shape fits averaged with 8 fits for the case of Fig. \ref{fig:Figure3}d-f is $\pm 9 \times 10^{-4}$ $cm^{-1}$ (30 MHz). For an InAs QD with $5 \times 10^4$ atoms, one arrives at a nuclear spin sensitivity of $\sim45$ nuclear spins. We believe that this very high precision is unprecedented in the field of optically-detected DNP studies of QDs under resonant excitation, resulting in fine-grained measurements of changes in nuclear spin ensemble. 

\begin{figure}[t]
  \includegraphics[width=0.4\textwidth]{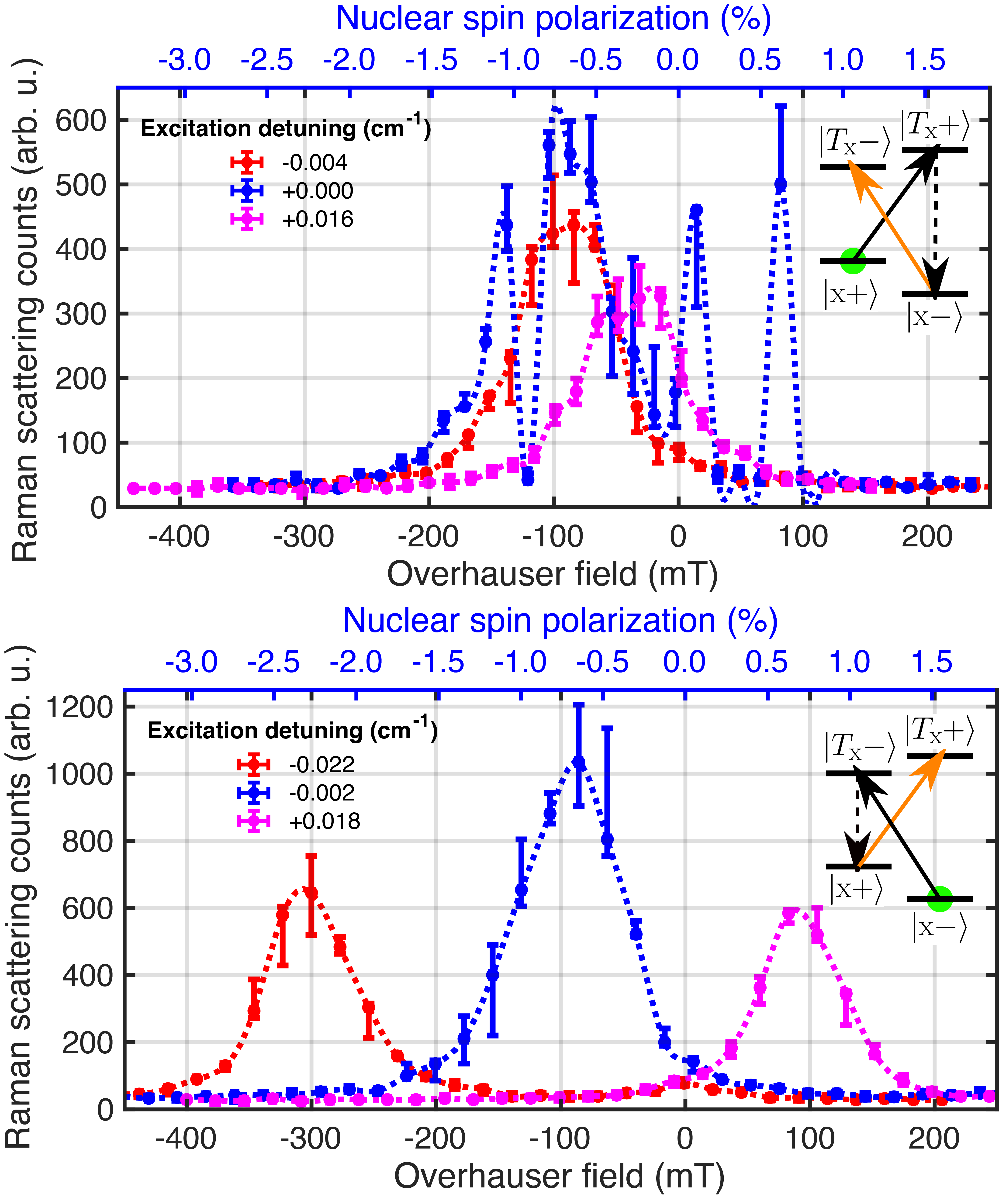}
  \centering
  \caption{OH field distributions extracted from the excitation-emission maps, with data collected in the optical pumping regime. (a) Excitation (solid black) (re-pump, solid orange) laser is backward-scanned across (held in resonance with) the $\ket{x+} \rightarrow \ket{T_x+}$ ($\ket{x-} \rightarrow \ket{T_x-}$⟩) transition, and the anti-Stokes Raman scattering (dashed arrow) is frequency-resolved. The electron spin is predominantly prepared in the $\ket{x+}$ state. (b) Excitation (re-pump) laser is forward-scanned across (held in resonance with) the $\ket{x-} \rightarrow \ket{T_x-}$ ($\ket{x+} \rightarrow \ket{T_x+}$) transition, and the Stokes Raman scattering is frequency-resolved. The electron spin is predominantly prepared in the $\ket{x-}$ state. Error bars are given by the variation in Raman scattering counts for the two adjacent excitation energy steps ($\pm$0.0002 cm$^{-1}$, $\pm$ 
  1 mT) around each excitation point indicated in the legends. }
\label{fig:OHfield}
\end{figure}

In the electron spin-up $\ket{x+}$ case, there is a rapid switching of the NSP as the excitation laser frequency passed through the trion resonance in the $\ket{x+}$ case, compared to the smooth tracking of the OH field in the $\ket{x-}$ case. Near the trion resonance for the $\ket{x+}$ case, the OH field distribution becomes multi-modal, as evident in Fig. \ref{fig:OHfield}a. Away from the resonance, the nuclear polarization distribution is single-peaked; this data is collected from the optical pumping case in which the excitation laser is backwards-scanned across the $\ket{x+} \rightarrow \ket{T_x+}$ transition (Appendix \ref{additional}). The multi-modality of the distribution indicates an increase of optically-induced fluctuations of the OH field, revealing an instability point of the coupled electron-trion-nuclear system. The transition from a stable configuration, represented by a single-peaked distribution with a FWHM of approximately 90 mT, to an unstable point occurs within less than 0.012 cm$^{-1}$ (360 MHz). The widths of the narrow peaks observed at the instability point are less than the FWHM of the detection etalon, implying that the OH field changed by at least 75 mT between consecutive laser scans (repeated identical experiments). Further examples of observations of instability points are demonstrated by visual examination of Fig. 9c, where abrupt jumps in the Raman scattering intensity are observed along the emission detuning axis for a fixed excitation detuning. 

This data shows that, in the avoidance case, setting the excitation laser to certain detuning ranges especially near zero detuning leads to an increase in optically-induced fluctuations of the OH field, which map onto fluctuations in the Raman scattering intensity between adjacent detection energies. The timescales for these large OH field fluctuations are longer than the time taken for the laser to scan from one detuning to the adjacent detuning (25 ms) as indicated by the error bars in Fig. \ref{fig:OHfield}a, but shorter than the time taken to construct a single excitation energy scan (15-25 seconds, depending on scan range).

For comparison, in the case where the excitation laser is scanned across the $\ket{x-} \rightarrow \ket{T_x-}$ transition and the OH field tracks to hold the trion transition in resonance with the laser, the OH field distributions are single-peaked (Fig. \ref{fig:OHfield}b), and do not exhibit any instability points at which the OH field fluctuates between consecutive laser scans by more than expected for a Gaussian distribution with 90 mT FWHM. 

The implications of these findings regarding the OH field distributions are significant. First, instability points have been predicted theoretically to result from the non-linear feedback loop between the OH field acting on the electron and the electron polarization and/or trion populations acting on the nuclei, which are tuned by the scanning excitation laser; these instability points can correspond to points of increased OH field distribution width \cite{Yang2012,Yang2013,Onur2018}. At these points, there is an amplification, or increase, of the fluctuations of the OH field; we have confirmed these predictions here experimentally in the case of dynamic optical excitation (scanning the excitation laser) of the $\ket{x+}$ electron spin state. 

Second, there clearly exist stable points of the electron-trion-nuclear system, especially when the laser scans across the $\ket{x-} \rightarrow \ket{T_x-}$ transitions, represented by the single-peaked OH field distributions, and larger OH fields which are reinforced through strong negative feedback loops which lock the OH field in place, as predicted by Yang and Sham\cite{Yang2012,Yang2013}. That theoretical work implies that the OH field distributions are narrowed in this excitation regime. Previous experimental studies have focused on the narrowing effect induced by coherent population trapping (CPT) of the electron spin \cite{Xu2009,Sun2012a,Onur2016,Ethier-Majcher2017,Gangloff2018}. While we do not observe significant narrowing of the OH field as limited by our detection system, we do measure directly a relatively quiescent OH field in the tracking case compared to the avoidance case. 

Furthermore, the multi-modality of the OH field distribution may indicate the possibility of ordered nuclear spin clusters with varying fluctuations resulting from the nuclear dipole interaction \cite{Alvarez2015} as well as the electron-mediated nuclear-nuclear interaction \cite{Deng2005,Huang2010a,Latta2011}. These experiments thereby indicate that preparation of the electron spin in the $\ket{x-}$ state may be preferential for the purposes of quantum computation protocols: the OH fields measured in that configuration are observed to be relatively stable in comparison to the $\ket{x+}$ state.

The dynamics associated with OH field switching are in order of magnitude agreement with at least two previous measurements \cite{Maletinsky2007,Sun2012a}. In this experiment, rapid switching between OH field states is observed primarily in the ``avoidance'' regime: when the excitation laser is brought into resonance with the expected trion transition, the OH field switches by $\pm$ 150 mT within 25 ms. For comparison, one study measured the build-up time using pump/probe spectroscopy under the application of an out-of-plane magnetic field of -220 mT, with a result that varied between 11.2 and 32.3 ms \cite{Maletinsky2007}. Another study measured the polarization build-up time using CPT in the negatively-charged trion system to be equal to 7 $\pm$ 1 ms at 5 K under the application of an in-plane magnetic field of 2.64 T \cite{Sun2012a}.

\section{Microscopic mechanisms leading to the generation of nuclear spin polarization}

\begin{figure}[t]
  \includegraphics[width=0.5\textwidth]{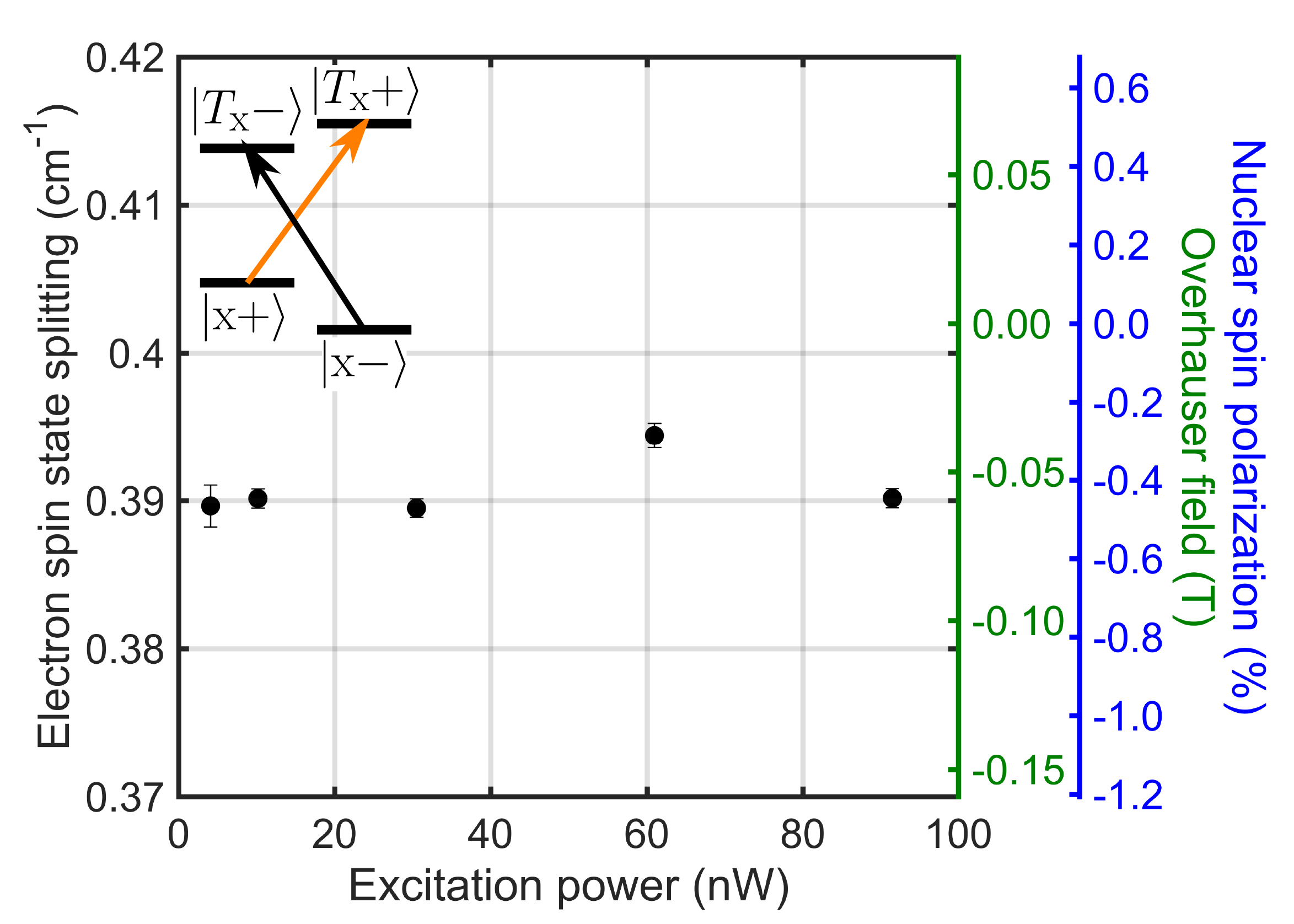}
  \centering
  \caption{Excitation/re-pump power ratio Raman scattering experiment. The re-pump(orange) (excitation, black) is held on resonance with the $\ket{x+}\rightarrow\ket{T_x+}(\ket{x-}\rightarrow\ket{T_x-})$ transition, and the etalon detection energy is scanned across the four scattering lines. The excitation power is changed while the pump power is held constant at 20 nW for a given spectrum. The saturation power using the excitation field is approximately four times higher than the re-pump due to intentional misalignment. The energy axes are chosen to approximately match that of Figure 3(c,f) for meaningful physical comparison. }
\label{fig:Figure4}
\end{figure}

Various microscopic mechanisms have been proposed to account for the DNP effects observed in InAs QDs. These mechanisms include the electron-nuclear hyperfine contact interaction \cite{Meier1984,Abragam1961,Slichter1992,Gammon2001}, the hole-nuclear dipole interaction \cite{Koudinov2004,Fischer}, and the strain-induced quadrupole-mediated electron hyperfine interaction \cite{Latta2011,Bulutay2012,Chekhovich2012a}, as well as additional depolarization terms \cite{Maletinsky2007}.

The electron-nuclear Fermi contact interaction is manifest in QDs via two terms. First, recalling the applied magnetic field is in the x-direction (in-plane), the collinear term is given by the summation over the nuclear spin ensemble $\sum_j A_{e}^j I^{j}_x S_{x}$, directly leading to the OH field that shifts the electron spin splitting in the presence of a non-zero NSP. Second, the so-called flip-flop term, given by $\frac{1}{2}\sum_j A_{e}^j \left( I_{+}^j S_{-} + I^{j}_- S_{+} \right)$, allows for transfer of angular momentum from the electron to the nuclear spin ensemble \cite{Abragam1961}, and the balance between the two channels may be influenced by preferential optical excitation of one of the trion transitions. 

The relevance of the flip-flop interaction is experimentally investigated by measurement of the OH field in the presence of optical preparation of an electron spin polarization. In the experiment (Fig. \ref{fig:Figure4}), which is operated in the optical pumping regime, the re-pump (excitation) laser is fixed in resonance with the $\ket{x+}\rightarrow\ket{T_x+}$ ($\ket{x-}\rightarrow\ket{T_x-}$) transition. The electron spin polarization is tuned by adjusting the excitation:re-pump power ratio between 1:5 and 22.5:5. We note that the  excitation and re-pump beams were spatially misaligned in order to achieve higher spatial rejection at the single-mode fiber before the spectral filtering stage (see Appendix E for more details). Raman scattering intensity experiments compared to a rate equation model showed that setting the excitation beam intensity 3-4x larger than the re-pump beam intensity initialized the electron spin state to equal parts $\ket{x\pm}$. Additionally,  the etalon is scanned across the four trion scattering lines, from which the electron spin splitting is extracted. 

The results indicate that no discernible OH shift occurs as a function of the power ratio, and by extension the electron spin polarization. Additionally, no increase in linewidth of the scattering is observed, indicating no change in the nuclear spin fluctuations. This experiment rules out the electron-nuclear flip-flop interaction as the primary mechanism for the OH feedback mechanism reported here. This finding is not surprising since this mechanism is inhibited by the mismatch between the nuclear spin Zeeman energy at 2 T (75 neV) and a number of other parameters in the system, including the electron Zeeman splitting ($\sim\text{ }50\mu\text{eV}$), and the 1LO phonon energy (37 meV) \cite{Urbaszek2013}, as well as weak coupling of the electron-trion system to acoustic phonons \cite{Hansom2014}. 

%%% HHLH mixing discussion

Another microscopic mechanism under investigation, the heavy-hole light-hole mixing term (HHLH), is strongest in anisotropic QDs \cite{Pikus1994,Leger2007,Koudinov2004,Krizhanovskii2005,Bulaev2005,Fischer,Eble2009,Xu2009,Belhadj2010,DeGreve2011, Greilich2009,Sun2012a, Carter2014, Huthmacher2018}. In such QDs, the mixing of spatial wavefunctions of the optically excited holes in the valence band enables the hole-nuclear dipole interaction. Even in the absence of HHLH mixing, this interaction is present, although considerably weaker than the Fermi hyperfine contact term \cite{Fischer}. HHLH mixing manifests as an optically measurable quantity via rotation of the optical selection rules \cite{Leger2007,Testelin2009}. Sample bias-modulated polarization-sensitive reflectivity measurements reveal that the selection rules are rotated away from the external magnetic field axis by no more than $(8\pm3)^{\circ}$ (Appendix \ref{HHLH}); for comparison, a previous experimental study that implicated the HHLH mixing term as the primary mechanism for DNP measured a rotation of the selection rules by $20^{\circ}$ \cite{Xu2009}. Thus, it is concluded that the HHLH mixing-induced hole-nuclear dipole term is less relevant in this QD than the strain-induced quadrupolar term (described below), although a direct comparison of the strength of each term is difficult.

%%% quadrupolar mechanism

The other primary term in the Hamiltonian that can generate NSP in the QD system is the quadrupole-mediated electron hyperfine interaction, which plays a role in highly strained QDs fabricated using the Stranski-Krastonov growth technique \cite{Chekhovich2015}. Non-zero electric field gradients present in the strained crystal lattice lead to a mixing of nuclear spin states \cite{Abragam1961,Chekhovich2015,Chekhovich2012a}; perturbation theory leads to a Hamiltonian term of the form $H_{quad}^{elec} = \sum_i A_{nc} S_x^e \left[ I^i_+ + I^i_- \right]$, which has previously been incorporated into models in order to explain optically-induced NSP effects in QDs. \cite{Issler2010,Hogele2012a,Schrieffer1966}. NSP may be generated in the QD without flipping the electron spin, thereby enabling an energetically-allowed polarizing channel. Additionally, the lack of dependence on the heavy-hole state points towards the role of an electronic effect. 

The origin of the electron-nuclear spin coupling arises due to strain in the QD system. Here we address the effects of strain and in-plane anisotropy in the QD as related to the quadrupolar interaction, HHLH mixing, and the observation of non-zero in-plane hole g-factors. It has been noted previously that in-plane anisotropy, leading to the reduction of symmetry in the QD away from D$_{2d}$ (x = y), results in elliptically-polarized selection rules in the trion, as well as non-zero in-plane g-factors for the negative trion \cite{Belykh2016,Koudinov2004,Belhadj2010}. This reduction may manifest via shape anisotropy or in-plane strain. However, it has also been shown theoretically that even in D$_{2d}$ symmetry  InAs/GaAs QDs that do not exhibit HHLH mixing that non-zero in-plane g-factors result due to strong 0D quantization which quenches orbital angular momentum \cite{Pryor2006}; in that case, one possible situation arises in which the hole and electron have opposite sign in-plane g-factors, as observed here.

Additionally, it is noted that the quadrupolar interaction is a local effect in that strain, whether along the growth axis or in-plane, yields crystal fields which cause the mixing of individual nuclear spin states due to a local re-orientation of principal axes \cite{Bulutay2012,Bulutay2014}. No QD in-plane anisotropy as would be measured by HHLH mixing is required for strong quadrupolar interactions. However, the quadrupole interaction is not directly manifested in an optically measurable quantity such as a selection rule rotation. Thus, further theoretical work is required to confidently attribute the DNP phenomena observed here to the quadrupolar mechanism.

\section{Discussion}
Strong evidence is provided for two distinct non-linear feedback regimes of optically-induced NSP response in a single InAs QD using the single-photon resonant Raman scattering technique described above. The quantum-confined electron spin interacts with the collective NSP state, comprised of 10$^4$-10$^5$ nuclear spins, and induces an optical detuning via the OH field. In the case of scanning the laser across the $\ket{x-}$ electron spin state, the NSP responds to a dynamic excitation laser by tracking the laser: an OH field is generated that locks the trion transitions in step with the scanning laser over a wide excitation energy range. This feedback loop between the NSP and locking of the trion resonance to the laser results in a large and stable OH field, when compared to the $\ket{x+}$ electron spin case. This feedback regime is also marked by nuclear spin fluctuations well below 75 mT. This finding presents new insight into the stability of the NSP in the single InAs QD; quantum information protocols involving pure state initialization may benefit from initialization into the $\ket{x-}$ electron spin state due to reduced nuclear spin fluctuations and built-in robust protection against accidental laser detunings on the GHz level. Additionally, the Raman-scattered photons are highly tunable, and will find use as a photonic resource in quantum network schemes.

Excitation of the $\ket{x+}$ electron spin state stands in stark contrast to the excitation of the $\ket{x-}$ state; nuclear spin fluctuations are increased as the laser scans across the expected trion resonance, and the NSP exhibits multi-modality, switching on timescales faster than 25 ms between configurations that differ by more than 1.2\%. In this regime, these experiments have laid the groundwork for inducing fast changes in the nanoscopic state of the nuclear spin ensemble with exquisite sensitivity to the laser excitation energy. This result may prove useful for quantum sensing work, as well as fundamental investigations into many-body theories.

The NSP response to a dynamic laser excitation field was measured with unprecedented resolution down to the sub-100 nuclear spin level, as limited only by the etalon linewidth, allowing us to investigate fundamental physics problems such as the central spin problem. This study stands in contrast to previous measurements of optically-induced DNP under resonant excitation of InAs QDs. The determination of the OH field, and therefore NSP, is performed by direct measurement of the energies of the Raman- and Rayleigh-scattered photons, rather than ambiguously from absorption measurements. 

The utilization of a higher finesse etalon combined with deconvolution techniques should allow not only direct measurements of the electron spin coherence via the Raman scattering linewidth, but may also \textit{approach the single nuclear spin sensitivity limit}. Investigations of the NSP states at the single nuclear spin level will certainly provide insight into problems of quantum magnetism such as frustrated magnetism and quantum phase transitions. These studies are essential for the utilization of the nuclear spin ensemble as a quantum memory, further demonstrating the potential of the QD electron-trion-nuclear system in the quantum computing toolbox.

\begin{acknowledgments}
A.M.R. and D.G.S. acknowledge support from NSF Grant 1708062. L.J.S. acknowledges support from NSF Grant 1707970. D.G., A.S.B., and M.K.Y. acknowledge support from the Office of Naval Research. 
\end{acknowledgments}

\appendix

\section{Polarization-sensitive reflectivity measurements of g-factors and HHLH mixing}
\subsection{g-factor measurements} \label{gfactor}

\begin{figure}[h]
  \includegraphics[width=0.4\textwidth]{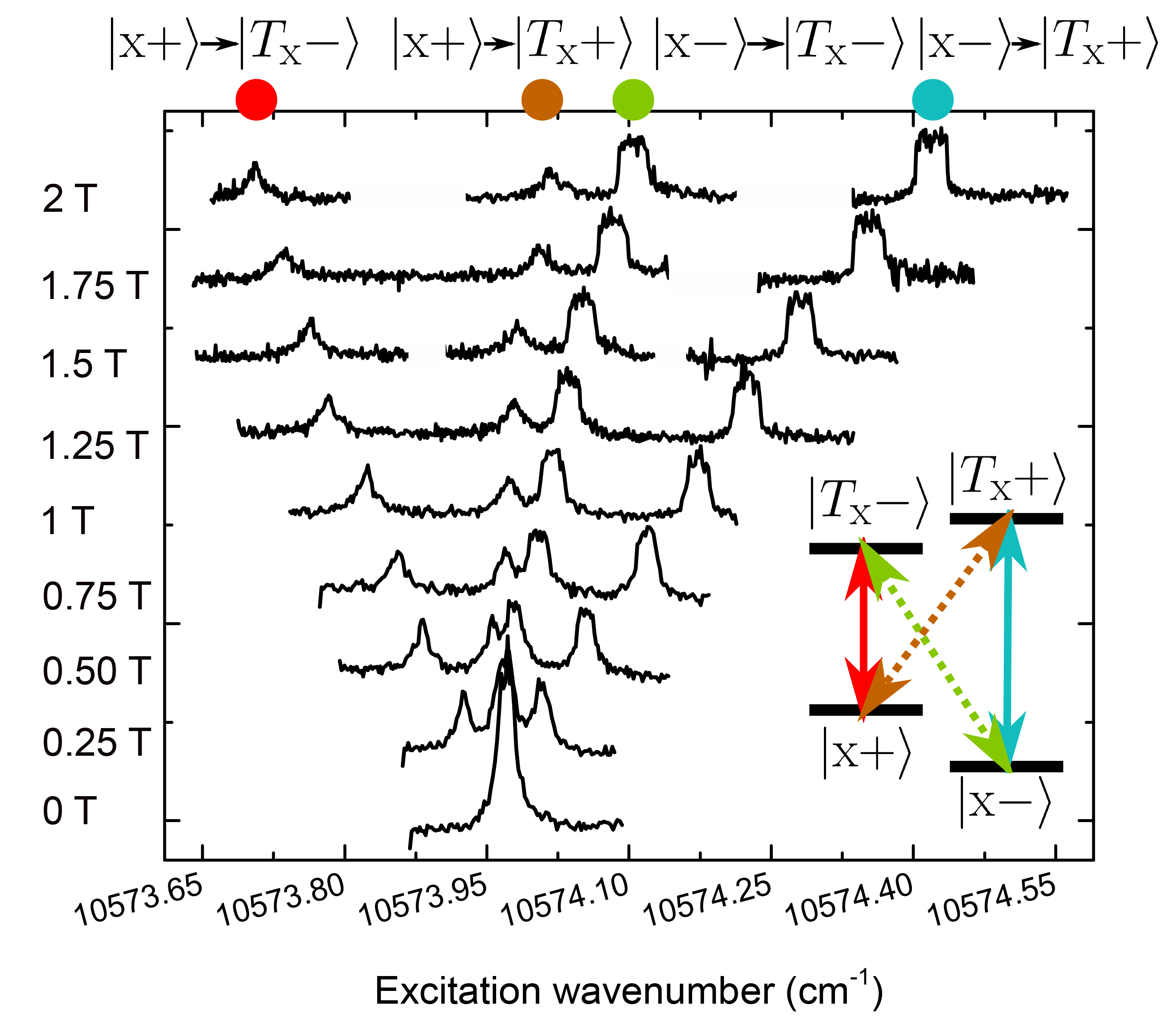}
  \centering
  \caption{Fan diagram identifying the electron-trion system. A resonant excitation laser is scanned across all four trion transitions and the modulated reflectivity is measured, where the magnetic field is increased in 250 mT steps for each trace. Inset: energy level diagram including selection rules and color-labelled transitions. Colored dots below each resonance match the corresponding color in the energy level diagram. Solid (dashed) lines in the energy level diagram correspond to horizontally (vertically) polarized selection rules. Transitions between ground and excited states are labelled above the plot.}
\label{fig:FanDiagram}
\end{figure}

To identify the trion, an in-plane magnetic field (Voigt geometry) is applied, taking advantage of the Zeeman effect to split the trion into four linearly-polarized resonances. For the QD under study, the four absorption spectral lines are clearly resolved beyond their linewidths for magnetic fields greater than 1 T; the Raman scattering and pump-probe experiments are performed at 2 T (Figure \ref{fig:FanDiagram}). The sample is operated in the co-tunneling regime to allow for single beam excitation without optical pumping of the electron ground states.

The lineshapes deviate significantly from the expected Lorentzian lineshapes; the two red Zeeman transitions (two lowest energy transitions) which correspond to scanning the resonant laser over the $\ket{x+}\rightarrow\ket{T_x-}$(red dot in Figure \ref{fig:FanDiagram}) and $\ket{x+}\rightarrow\ket{T_x+}$(brown dot in Figure \ref{fig:FanDiagram}) transitions, have sharp cusps where a rounded Lorentzian is expected. The two blue Zeeman transitions, corresponding to the $\ket{x-}\rightarrow\ket{T_x-}$(green dot in Figure \ref{fig:FanDiagram}) transition and $\ket{x-}\rightarrow\ket{T_x+}$(blue dot in Figure \ref{fig:FanDiagram}) transition, are broadened significantly, with widths of around 4.3 $\mu$eV (0.035 cm$^{-1}$, 1.04 GHz) compared to the zero field linewidth of $\sim$500 MHz, with flat tops across the broadened range over which the reflectivity signal is approximately constant. 

\begin{figure}[t]
  \includegraphics[width=0.4\textwidth]{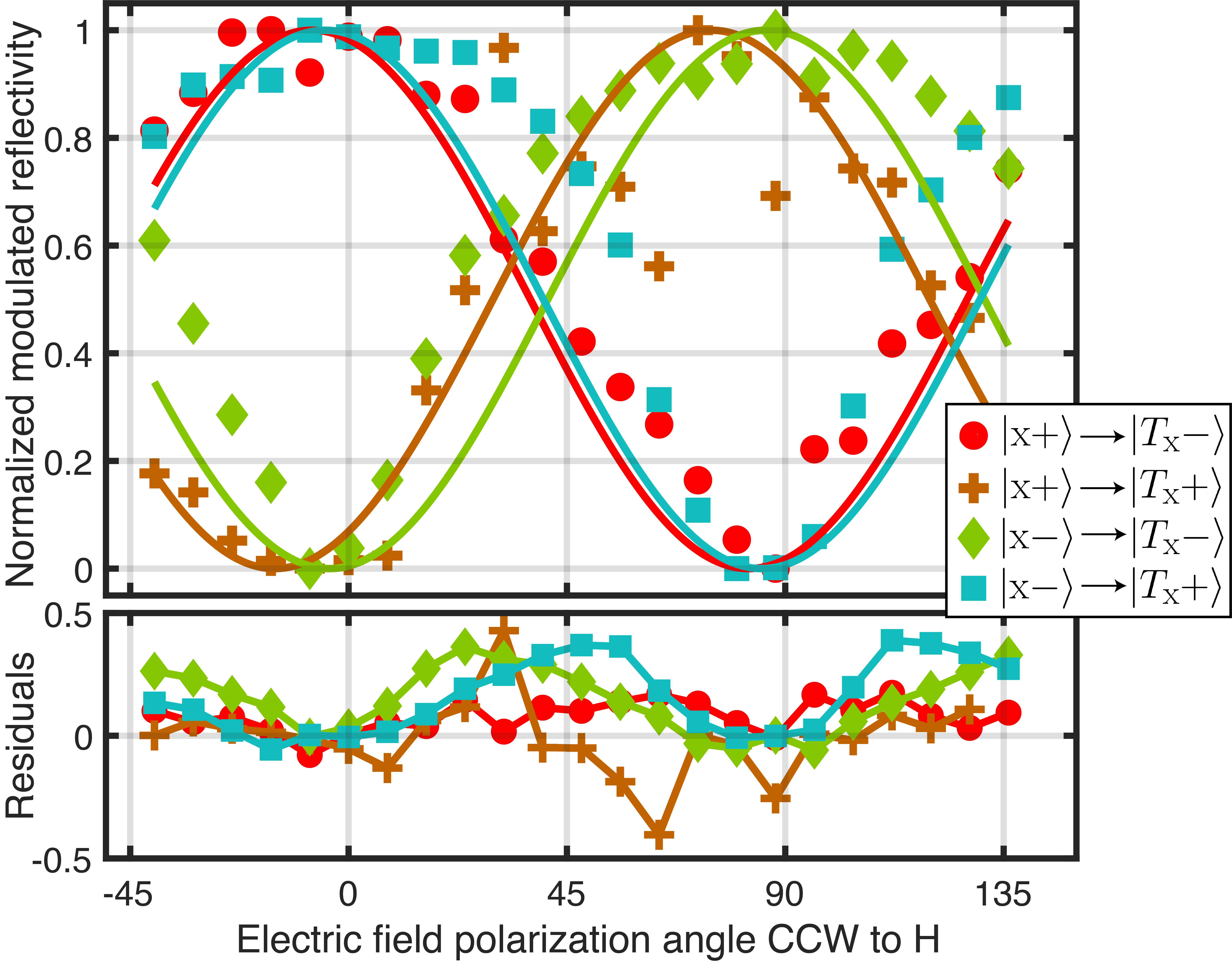}
  \centering
  \caption{Polarization selection rules of the QD trion transitions measured by polarization-sensitive lock-in reflectivity spectroscopy. Top: Markers are collected data points, solid curves are fits. Each curve is color-coded corresponding to its resonant electron-trion transition. Horizontal (H) means along the magnetic field direction $\hat{x}$, with the angle being measured counter-clockwise (CCW) in-plane relative to $\hat{x}$. Bottom: fit residuals.}
\label{fig:HHLHmixing}
\end{figure}

The magnitude of the in-plane electron and heavy-hole g-factors is determined by fitting the fan diagram (Figure \ref{fig:FanDiagram}).  Under these conditions, measurement of the g-factors assumes that large OH field shifts are eliminated by rapid depolarization due to tunneling electron spins between the QD and the reservoir, and the resonance center is determined as the centroid of the reflectivity peak \cite{Maletinsky2007}. Although there is DNP broadening present in the co-tunneling spectra that were used to measure the g-factors reported above, there is very little hysteresis ($<$ 60 MHz) observed in these spectra, implying that the underlying bare (free of DNP) trion resonances sit at the center of the co-tunneling resonances. 

The Zeeman Hamiltonian is taken to be equal to
\begin{equation}
    H_{Zeeman} = \mu_B g_e^x S_e^x B_x - \mu_B g_{hh}^x J_{hh}^x B_x
\end{equation}
and the in-plane electron and heavy-hole g-factors are determined to be $g_e^x = 0.431 \pm 0.004$ and $g_{hh}^x = -0.332 \pm 0.004$, resulting in Zeeman shifts of $(6.00 \pm 0.05)$ GHz/T and $(4.64 \pm 0.05)$ GHz/T for the electron and heavy-hole, respectively. The g-factors are determined to have opposite signs based on the measured polarization selection rules; the in-plane electron spin g-factor has been chosen to be positive, in agreement with theoretical work \cite{Pryor2006} and previous convention\cite{Xu2007, Xu2009}. Determining the absolute sign of the g-factors goes beyond the scope of this work.

\subsection{HHLH mixing measurement} \label{HHLH}

The heavy-hole-light-hole mixing (HHLH mixing) of the QD trion under investigation in the main body of the paper was measured using polarization-sensitive reflectivity measurements (Figure \ref{fig:HHLHmixing}). Operating at co-tunneling, an excitation laser, intensity-locked to less than 5\%, was sent through a linear polarizer and then through a half-wave plate before being focused on to the QD, after which the reflected beam was collected onto an avalanche photodiode. The signal was detected using sample bias-modulated lock-in spectroscopy \cite{Alen2003}. In the presence of HHLH mixing, the linear selection rules are expected to deviate from the reference frame determined by the in-plane magnetic field direction $\hat{x}$, with the expected absorption strength given by $L(\theta) = c(a^2 + b^2 - 2 ab \cos 2(\theta+\phi))$, where $\theta, \phi$ are the polarizer angle and HHLH mixing angle, and the other parameters are fit parameters related to the HHLH mixing terms \cite{Belhadj2010}. Very little HHLH mixing is observed: by fitting the data in Figure \ref{fig:HHLHmixing}, an average deviation from the magnetic field frame of reference of $(8\pm3)^{\circ}$. This small deviation of the selection rules away from the magnetic field frame may potentially be explained by a small, unintentional rotation of the sample in the sample holder. Additionally, the presence of periodic residuals in the tracking case (blue and green curves in Figure \ref{fig:HHLHmixing}) may indicate polarization-dependent DNP effects that should be investigated in the future.

\begin{figure}[t]
  \includegraphics[width=0.4\textwidth]{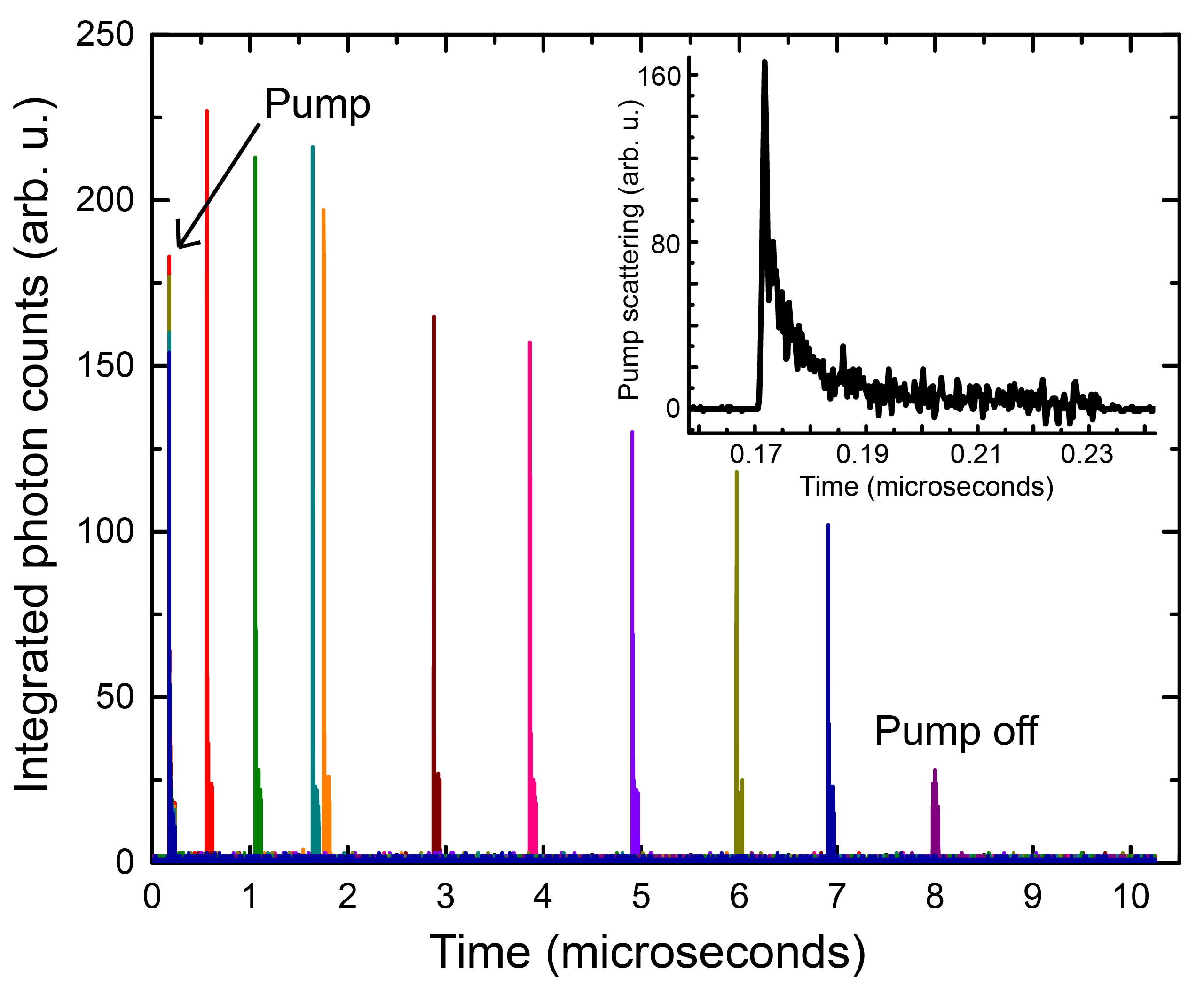}
  \centering
  \caption{Pump/probe delay experiment to measure $T_{1e}$. A series of TCSPC measurements for pump/probe delays varying between 400 ns and 7 $\mu$s. The pump(probe) drives the $\ket{x-}\rightarrow\ket{T_x-}$($\ket{x+}\rightarrow\ket{T_x+}$) transition. Different pump/probe delay TCSPC acquisitions are indicated by different color traces. The data are not background subtracted. }
  \label{fig:Delay}
\end{figure} 

\begin{figure*}[t]
  \includegraphics[width=0.95\textwidth]{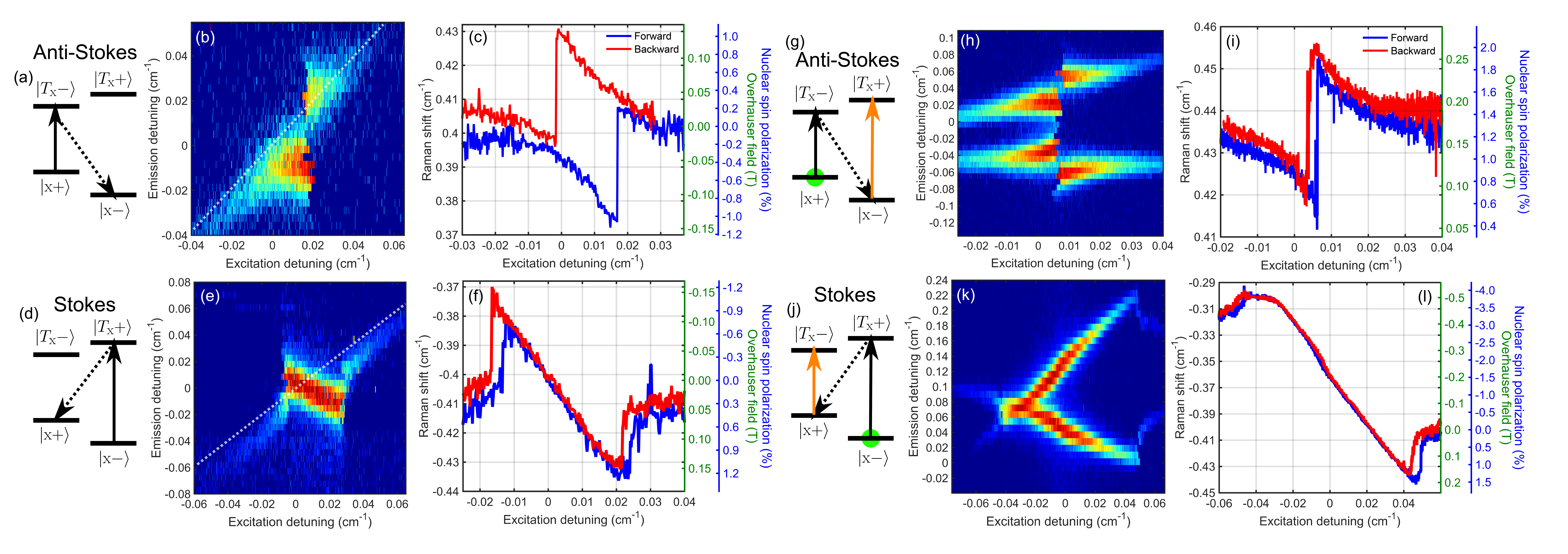}
  \centering
  \caption{Additional 2D Raman scattering excitation-emission energy maps and Raman shifts. (a-f): co-tunneling, (g-l): optical pumping. (a, g), (d, j): energy level diagrams for the anti-Stokes and Stokes scattering cases, with scanning excitation laser/re-pump laser (Raman scattering pathway) indicated with a solid black/solid orange (dashed) arrow, and the predominantly-prepared electron spin state indicated with a green circle. (b, h), (e, k): 2D Raman scattering maps corresponding to excitation schemes illustrated in (a, g) and (d, j), respectively. (c, i), (f, l): Raman shifts extracted by fitting each vertical cut of the 2D maps (b, h), (e, k) for a given excitation detuning, in both the increasing (blue) and decreasing (red) excitation laser energy directions.  Non-linear color coding adjusted as described below.}
\label{fig:OpticalSupplement}
\end{figure*}

\section{Time domain studies of optical pumping and electron spin relaxation} \label{timedomain}

In the main body of the paper it is asserted that the electron spin relaxation $T_{1e}$ is longer than 1 $\mu$s. Evidence is provided for this assertion by performing a time-domain pump/probe study using electro-optic modulators (EOMs, EOSpace) to optically pump the electron spin states using resonant excitation (Figure \ref{fig:Delay}). The two resonant pulses are approximately 60 nanoseconds, and the EOM contrast ratio is at least greater than 200. The pump(probe) drives the $\ket{x-}\rightarrow\ket{T_x-}$($\ket{x+}\rightarrow\ket{T_x+}$) transition. The pulse lengths are chosen to be long enough that the electron spin is optically pumped with high fidelity. The optical pumping fidelity is defined here as one minus the ratio of the counts at the end of the initialization pulse to the peak count rate at the beginning of the pulse as measured using time-correlated single-photon counting (TCSPC) (Figure \ref{fig:Delay}, inset). The initialization fidelity for the probe pulse is estimated to be equal to 94\%. The repetition period of the experiment, or the time between consecutive pump pulses, is set to 10.26 $\mu$s. The pump and probe scattering is measured using TCSPC via a HydraHarp module with the acquisition trigger synced to the pump EOM pulse generator (HP8082A). 

The delay between the pump and probe is tuned from approximately 400 ns to 7 $\mu$s to measure the $T_{1e}$ population relaxation time (or the longitudinal relaxation time in NMR language). No assumptions are made about the co-tunneling time but assume that it contributes to the population relaxation time measured here. Co-tunneling de-populates the QD of the electron and upon re-injection into the QD  drives the measured electron spin to an equal mixed state of $\ket{x+}$ and $\ket{x-}$. The probe signal, corresponding to the electron spin polarization, drops by a factor of around 1/2 between 400 ns and 7 $\mu$s delay times. For comparison, when the pump is blocked and only the probe drives the QD, the signal drops down to the noise floor of the experiment ($\sim$25 counts in Figure \ref{fig:Delay}). Thus, it is claimed that the $T_{1e}$ time is at least as long as 1 $\mu$s.

\section{Additional Raman scattering data} \label{additional}

In both the co-tunneling and optical pumping (Figure \ref{fig:OpticalSupplement}) cases, two more experiments were performed in which the other optically-excited heavy-hole state was driven. In the optical pumping cases, the scanning excitation laser transition was also switched, avoiding the coherent population trapping condition, and allowing for unambiguous assignment of a given scattering line to the re-pump and probe laser fields.

While slight differences in the quantitative responses of the DNP feedback are observed between these data and the data reported in the main body of the paper, no systematic qualitative differences are observed. Interpretation of the optical pumping data is complicated by the presence of two scattering lines, one from the pump and another from the scanning probe. However, it is relatively simple to separate these two lines in analysis, since the slope of Raman scattering dependence of the pump line should be flat far away from resonance (minimal DNP), compared to the scanning excitation laser dependence which follows the laser.

\section{Raman scattering map color scaling} \label{coloring}

The 2D excitation-emission maps are processed using a non-linear function of the raw photon counts to enhance the contrast between the signal and the noise. This processing function is applied to the maps after the Lorentzian lineshape fitting is performed and is only used to benefit the qualitative understanding of the OH field detuning in the different excitation/re-pumping cases. The processing algorithm is performed as follows:

\begin{enumerate}
\item $\text{Counts2} = \text{Counts1}-Min(\text{Counts1})$
\item $\text{Counts3} = \frac{\text{Counts2}}{Max(\text{Counts2})}$
\item If(Counts3(i)$<$ threshold): Counts4(i) = 0. Else: Counts4(i) = Counts3(i).
\item CountsFinal = arctan(stretch $\times$ Counts4)
\end{enumerate}

The two constants `stretch' and `threshold' are chosen arbitrarily for each map to enhance contrast. Unprocessed data may be provided upon request.

\section{Remarks about intentional mis-alignment of re-pumping beam}  \label{realign}
In order to enhance rejection of the pump beam at the collection optical fiber, the following procedure was used. Initially, resonant optical pumping experiments are performed that require both lasers to yield an absorption signal. Then, in order to increase spatial rejection of the re-pump laser before the fiber to the etalon/detection stage, the re-pump laser is re-aligned iteratively so as to increase the lateral distance between the two beams in both the excitation and collection paths, while retaining an optical pumping signal. During this process, increasing the off-axis distance of the re-pump beams induces an astigmatism, which reduces the optical pumping signal for a given re-pump power, since the beam spot is larger at the QD. Nevertheless, we ensure that the QD is still excited by the beam while enhancing re-pump spatial rejection before the collection fiber. While enhancing the spatial rejection of the pump beam and improving the quality of Rayleigh scattering lines, this process makes it difficult to compare excitation/re-pumping beam powers, especially for interpreting the results presented in Figure 5. 

% \bibliography{DNP}

%apsrev4-2.bst 2018-12-27 (MD) hand-edited version of apsrev4-1.bst
%Control: key (0)
%Control: author (8) initials jnrlst
%Control: editor formatted (1) identically to author
%Control: production of article title (0) allowed
%Control: page (0) single
%Control: year (1) truncated
%Control: production of eprint (1) enabled
%

\end{document}